\documentclass[
aps,prb,twocolumn,showpacs,superscriptaddress,citeautoscript,reprint]{revtex4-1}

\usepackage{amsmath}
\usepackage{amssymb}
\usepackage{graphicx}
\usepackage{units}
\usepackage{bm}
\usepackage{xcolor}
\usepackage[colorlinks]{hyperref}

\newcommand{\vc}[1]{\bm{#1}}

\newcommand{\K}{{\vc K}}
\newcommand{\vF}{{v_{\rm F}}} 
\newcommand{\AB}{{A_{\rm B}}} 

\newcommand{\rmi}{{\rm i}}

\begin{document}

\title{Spin-dependent THz oscillator based on hybrid graphene superlattices}

\author{E.\ D\'{\i}az}
\author{K.\ Miralles}
\author{F. Dom\'{\i}nguez-Adame}

\affiliation{\mbox{GISC, Departamento F\'{\i}sica de Materiales, 
Universidad Complutense, E-28040 Madrid, Spain}}

\author{C. Gaul}
\email{cgaul@pks.mpg.de}
\affiliation{Max Planck Institute for the Physics of Complex Systems, 
\mbox{01187 Dresden,~Germany}}


\begin{abstract}

We theoretically study the occurrence of Bloch oscillations in biased hybrid 
graphene systems with spin-dependent superlattices. The spin-dependent 
potential is realized by a set of ferromagnetic insulator strips deposited 
on top of a gapped graphene nanoribbon, which induce a proximity exchange 
splitting of the electronic states in the graphene monolayer. We numerically 
solve the Dirac equation and study Bloch oscillations in the lowest 
conduction band of the spin-dependent superlattice. While the Bloch 
frequency is the same for both spins, we find the Bloch amplitude to be spin 
dependent. This difference results in a spin-polarized ac electric current 
in the THz range.

\end{abstract}

\maketitle

According to the theoretical work by Esaki and Tsu,\cite{Esaki70} negative 
differential resistance in biased superlattices~(SLs) signals the occurrence 
of Bloch oscillations~(BOs).\cite{Bloch29,Zener34} One of the main obstacles 
for the realization of an active Bloch oscillator is the instability of 
the electric field, which results in the formation of electric domains. 
Savvidis \emph{et al.}\ found that these domains appear to be 
suppressed in a InAs/AlSb super-SL composed of many very short 
segments of SL, interrupted by heavily doped InAs regions.\cite{Savvidis04}
The stabilization of the electric field in semiconductor 
SLs can also be achieved by application of the cleaved-edge 
overgrowth technique.\cite{Feil05} 
In any way, the suppression of the electric 
domains in semiconductor SLs requires fairly complex designs.

Graphene SLs may easily overcome the instability of the electric field since 
the back gate voltage induces a uniform population of the quantum wells. In 
addition, the carrier density can be varied over a wide range. 
Patterning graphene at the nanometer scale can be achieved by hydrocarbon 
lithography\cite{Meyer08}, chemical functionalization,\cite{Sun11} or He ion 
lithography,\cite{Archanjo14} which opens a possibility to fabricate these SLs.
Graphene SLs have recently been a focus of interest to study a variety of 
quantum phenomena.\cite{Maksimova12,Dubey13,Ponomarenko13,Sattari13}
Dragoman and Dragoman proposed a SL obtained by patterning an array of metallic electrodes on gapless graphene,
where BOs of up to tens of terahertz 
can be generated due to the low scattering rate in graphene.\cite{Dragoman08}
In their design the metallic electrodes are inclined with respect to the current flow to 
minimize Klein tunneling. Negative differential resistance and the Wannier-Stark 
ladder regime in semiconducting armchair graphene nanoribbon~(GNR) SLs has 
been investigated by Ferreira \emph{et al.}\cite{Ferreira11} 
When the gap of the GNR is small, besides conventional BOs, multiple Zener tunneling between 
the coupled electron and hole branches leads to distinct coherent 
oscillations.\cite{Krueckl12}

Spin-related and magnetic effects are of special interest for their 
relevance in spintronics.\cite{Jiang11,Huo12,Lu12,Yu12,Faizabadi12} 
Recently, we proposed a hybrid SL realized by EuO ferromagnetic insulator 
strips deposited on top of a GNR.\cite{Munarriz13} 
These strips induce a proximity exchange splitting of the electronic states in 
graphene,\cite{Haugen08} resulting in the appearance of a SL with a 
spin-dependent potential profile. The electric current through the hybrid SL 
can be highly polarized and manifests spin-dependent negative differential 
resistance.\cite{Munarriz13} 

In this paper we investigate the high-frequency dynamics of electrons in a 
hybrid SL formed by a periodic arrangement of ferromagnetic 
strips grown on top of an armchair GNR. As mentioned above, the 
ferromagnetic strips induce a spin-dependent potential. Therefore, we 
expect spin-dependent BOs when the hybrid SL is subjected to a 
voltage drop between the source and the drain. We find that the Bloch 
frequency agrees with the semiclassical prediction and that the Bloch 
amplitude does so for sufficiently wide wave packets. Interestingly, as it 
occurs in the case of Bloch oscillators based on semiconductors, the 
present design also generates electric currents in the THz range. However, 
the amplitude of the BOs are spin dependent and consequently, the 
generated ac electric current is spin polarized. 

\begin{figure}[bt]
\includegraphics[width=\columnwidth,clip=]{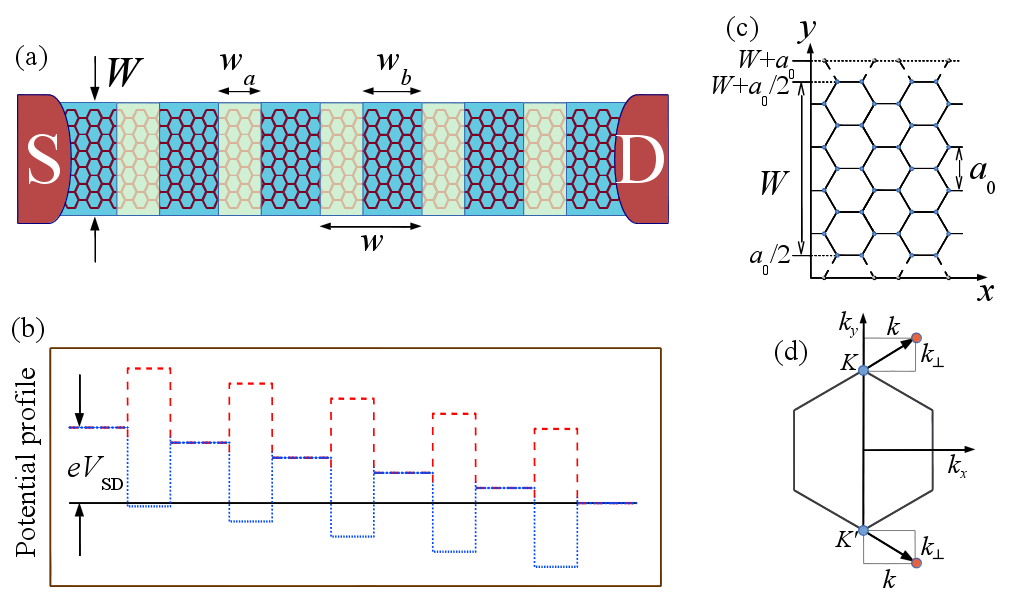}
\caption{
(a)~Sketch of the setup: The GNR is connected to source~(S) and drain~(D) 
leads, with $N=5$ strips of a ferromagnetic insulator on top of it. 
(b)~Potential profiles for spin-up (dashed red lines) and spin-down (dotted 
blue lines) electrons in the biased device.
(c)~Honeycomb lattice structure and edges of the GNR.
(d)~Brillouin zone of the honeycomb lattice with the Dirac points 
$\vc{K}$ and $\vc{K}'$ (blue) and two states (red) whose superposition 
fulfills the boundary conditions if $k_\perp$ is chosen correctly.
\label{figSetup}}
\end{figure}

The hybrid system consists of a rectangular GNR of width $W$,  connected 
to source and drain leads, on top of which there are $N$ ferromagnetic 
insulator strips of width $w_a$, with the spacing between them being $w_b$ 
[see Fig.~\ref{figSetup}(a)]. We restrict ourselves to GNRs with armchair 
edges hereafter. Experimental evidences\cite{Han07} and 
\emph{ab-initio\/} calculations\cite{Son06} show that the energy spectrum 
presents a gap inversely proportional to the width $W$, depending on the 
remainder $(2 W/a_0 \mod 3)$, where $a_0 = \unit[0.246]{nm}$ is the lattice 
constant, namely the width of the graphene lattice hexagon. Contrary to GNRs 
with zigzag edges, the dispersion relation of the armchair ones is centered 
around $k = 0$, making the resonant levels broader and less affected by 
disorder.\cite{Munarriz11}

The proximity exchange interaction between magnetic ions in the strips and
charge carriers in the GNR can be described as an effective Zeeman
splitting $\pm\Delta_\mathrm{ex}$ of the spin sublevels.\cite{Haugen08}
We use $\Delta_\mathrm{ex}=\unit[8]{meV}$ as a typical value; we have checked
that our results do not change qualitatively if we use a different value of
$\Delta_\mathrm{ex}$ within the range of a few $\unit{meV}$.
The proximity exchange interaction has the characteristic length scale of one
atomic layer. Therefore, the splitting is induced only in the  regions of the
GNR directly below the ferromagnetic strips. 
Consequently,  a spin-up (spin-down) electron propagating along the sample will
be subjected to a potential comprising a periodic set of rectangular barriers
(wells), as plotted in Fig.~\ref{figSetup}(b).

For not too narrow GNRs, the low energy excitations can be treated very
efficiently within the Dirac approximation.\cite{CastroNeto09} 
The wave function on each sublattice $\alpha = a,b$ is expanded around both 
Dirac points $\vc{K}$ and $\vc{K}'=-\vc{K}$, which are also referred to 
as \emph{valleys\/},
\begin{align}\label{eqDiracExpansion}
 \phi_\alpha(\vc{r}) 
 = e^{ \rmi \K\! \cdot \vc{r}} \psi_{\alpha}^{+}(\vc{r}) 
 + e^{-\rmi \K\! \cdot \vc{r}} \psi_{\alpha}^{-}(\vc{r})\ .
\end{align}
In coordinates where $\vc{K}$ points lie in the $y$ direction 
[see Fig.~\ref{figSetup}(d)] the Dirac equation reads
\begin{align}\label{Dirac}
 \rmi \hbar\, \frac{\partial}{\partial t} 
 \begin{pmatrix}
  \psi_a^\pm \\
  \psi_b^\pm
 \end{pmatrix}
 =
 \begin{pmatrix}
  \mathcal{V}(x) & \vF(-\rmi \hat{p}_x \mp \hat p_y) \\
         \vF( \rmi \hat{p}_x \mp \hat p_y) & \mathcal{V}(x)
 \end{pmatrix}
 \begin{pmatrix}
  \psi_a^\pm \\
  \psi_b^\pm
 \end{pmatrix}.
\end{align}
Here, $\vF=3t_0 a_0/2=\unit[10^6]{m/s}$ is the Fermi velocity in graphene, with 
$t_0$ being the nearest-neighbor hopping energy in the honeycomb lattice.
In $y$ direction, the boundary conditions require the wave function to vanish 
on the (fictitious) sites just
outside the GNR, \emph{i.e.}, at $y=0$ and $y=W+a_0$, where the $y$ axis
is perpendicular to the direction of the GNR, whose lower edge is located
at $y=a_0/2$ [see Fig.~\ref{figSetup}(c)]. 
In the present case of armchair GNRs, boundary conditions are fulfilled by a 
superposition of
two states from different valleys with the same longitudinal wave function but 
with opposite transverse wave number:
$\psi_\alpha^\pm(x,y) = \exp(\pm \rmi k_\perp y) f_\alpha(x)$.
The possible values of $k_\perp$ depend crucially on the width $W$ of the 
GNR and on ($2W \mod 3 a_0$).\cite{Brey06,Wakabayashi09}
In the gapped cases, which are of interest here, the lowest value is
$k_\perp \approx \pi/3W$. The transverse part 
of $\phi_\alpha$ is a rapidly varying standing wave of the form $\sin[(K+k_\perp)y]$.
The longitudinal wavefunction $f_\alpha(x)$, 
on the contrary, varies smoothly. Its equation of motion is a one-dimensional 
Dirac equation 
\begin{align}\label{eq1DDirac}
 \rmi \hbar\, \frac{\partial}{\partial t} 
 \begin{pmatrix}
  f_a \\
  f_b
 \end{pmatrix}
 =
 \begin{pmatrix}
  \mathcal{V}(x) & \hbar\vF(-\partial_x - k_\perp) \\
         \hbar\vF( \partial_x - k_\perp) & \mathcal{V}(x)
 \end{pmatrix}
 \begin{pmatrix}
  f_a \\
  f_b
 \end{pmatrix} .
\end{align}

For a constant potential $\mathcal{V}$, the solution of the 
one-dimensional Dirac equation in terms of plane waves is straightforward. 
For piecewise potentials, the solutions can then be matched together by 
transfer-matrix techniques.\cite{Munarriz13} Due to nonlinear dependence 
on $E-\mathcal{V}(x)$, however, the analytical treatment is difficult. 
Here, we pursue a different approach and seek to diagonalize the 
stationary version of~\eqref{eq1DDirac}.

The Dirac equation~\eqref{eq1DDirac} couples $f_\alpha$ to the first derivative 
$f^{\prime}_{\bar\alpha}$ of the respective other sublattice. Thus, it is convenient 
to sample $f_b$ on points that lie just between the points where $f_a$ is sampled 
and to collect the data in an alternating array $g_j$ with 
$g_{2n} = f_a(2n d)$, $g_{2n+1} = f_b\big((2n+1)d\big)$ and discretization step $d$.
The stationary one-dimensional Dirac equation \eqref{eq1DDirac} then becomes 
a one-dimensional tight-binding equation of motion with alternating hopping 
energies
\begin{align}\label{eqDiscrete1DDirac}
\frac{E-\mathcal{V}_j}{\hbar\vF}\,g_j = (-1)^j \frac{g_{j+1}\!-\!g_{j-1}}{2d} 
-  k_\perp \frac{g_{j+1}\!+\!g_{j-1}}{2} \ ,
\end{align}
where $\mathcal{V}_j=\mathcal{V}(j d)$.
The last term comes from interpolating the wave function on the opposite 
sublattice for $f_b(2n) = [f_b((2n+1)d)\!+\!f_b((2n-1)d)]/{2}$ and similar 
for $f_a\big((2n+1)d\big)$. For vanishing potential, the spectrum 
of~\eqref{eqDiscrete1DDirac} as a function of the longitudinal momentum $k$ is
\begin{equation}\label{eqDiscreteSpectrum}
 E_\pm(k) = \pm \hbar \vF \left[ k_\perp^2 \cos^2(k d) + d^{-2} \sin^2(k d)\right]^{1/2} ,
\end{equation}
with a gap opened by the transverse momentum $k_\perp$. In the limit $d\to 
0$, this goes over to the well-known Dirac dispersion $E_\pm(k) = \pm 
\hbar \vF (k_\perp^2 + k^2)^{1/2}$. Eq.\ \eqref{eqDiscreteSpectrum} holds for low energies. 
The outer band edges $\pm \hbar\vF/d$ are artifacts of the discretization and have nothing to do 
with the band edges of the honeycomb lattice. 

\begin{figure}[bbt]
\includegraphics[clip,trim=20 15 10 5]{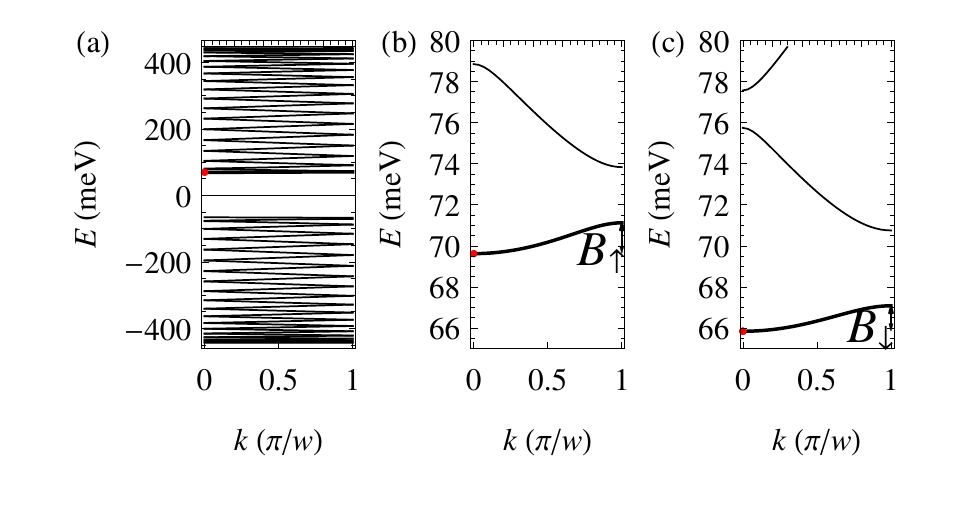}
\caption{Band structure from~\eqref{eqDiscrete1DDirac} in the case of an infinite 
untilted SL with $W=\unit[9.84]{nm}$, $\Delta_{\rm ex} = \unit[8]{meV}$, 
$w_a=\unit[23.9]{nm}$, $w_b=\unit[83.0]{nm}$, and discretization step
$d=w/72=\unit[1.48]{nm}$ with
$w=w_a+w_b$. (a)~Full spectrum for a spin-up electron. Panels~(b) 
and~(c) show enlarged views of the lowest conduction bands for spin up
and down, respectively.}
\label{figBandstructure}
\end{figure}

Figure~\ref{figBandstructure} presents the numerically obtained band structure 
for the infinite untilted SL.
While the central gap is again due to the transverse momentum 
$k_\perp$, the other gaps are due to the SL strength 
$\Delta_{\rm ex}$. Interestingly, the widths of the bands $B_\uparrow$ and
$B_\downarrow$ are different ($1.53$ and $\unit[1.26]{meV}$ for the
chosen parameters, respectively). Since the amplitude of the BOs depends on
the bandwidth, this difference will ultimately lead to the generation of 
spin-dependent ac electric current.

We are now in position to numerically diagonalize the Dirac equation. 
Keeping in mind the properties of the untilted SL, $d$ has 
to be sufficiently small, such that the SL strength $\Delta_{\rm ex}$ 
is smaller than the bandwidth~\eqref{eqDiscreteSpectrum} to avoid 
discretization artifacts, namely $d \ll \hbar \vF/\Delta_{\rm ex}$. 
Hereafter we take $d=w/72$ with $w=w_a+w_b$. We have checked that the results 
are the same within the numerical uncertainty for smaller values of the 
discretization step. Moreover, in order to have electron and hole 
states well separated, the SL strength should not exceed the gap 
of the homogeneous GNR, i.e., $\Delta_{\rm ex} \ll \hbar \vF k_\perp$.

\begin{figure}[tb]
\includegraphics[width=0.75\columnwidth,clip]{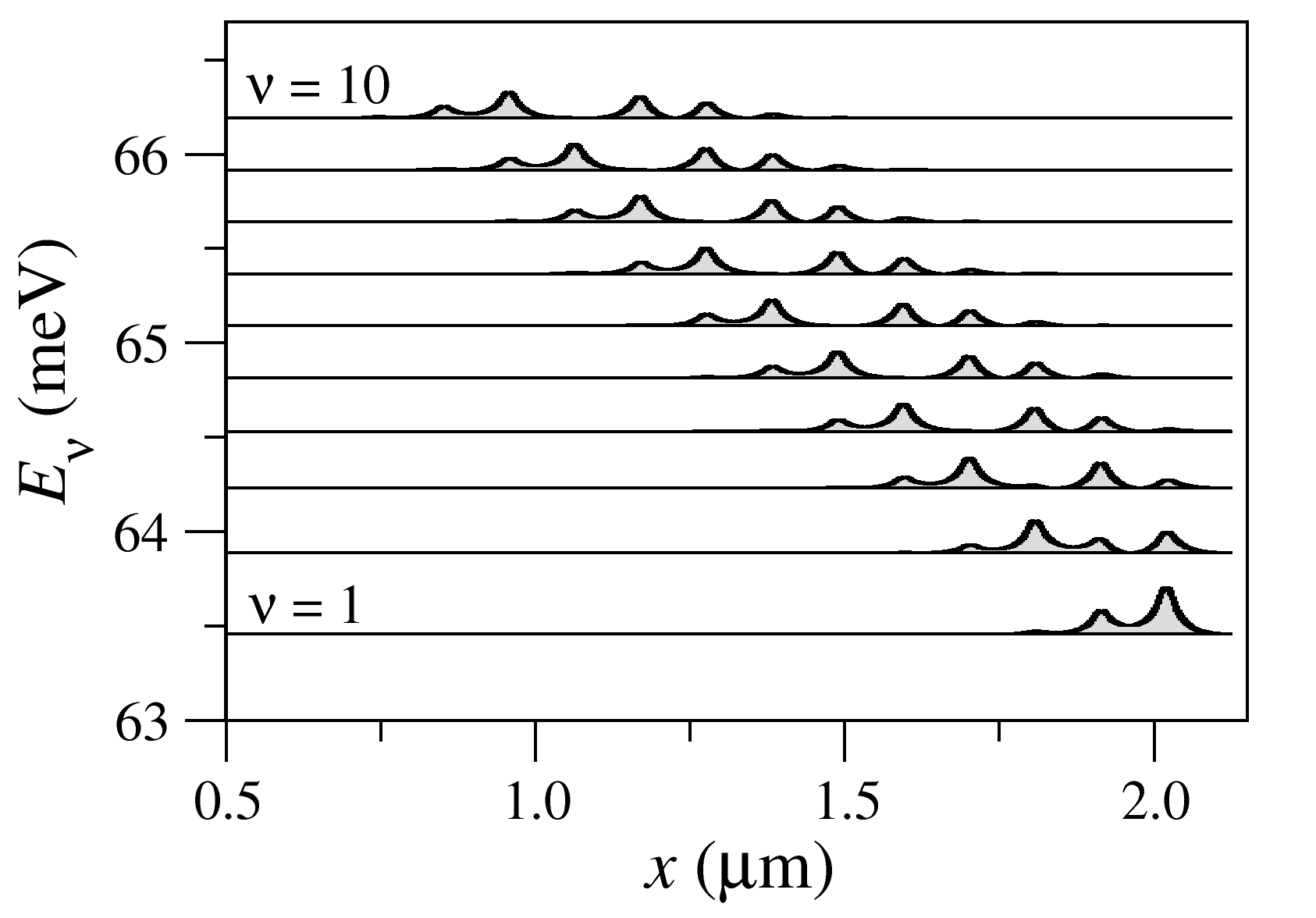}
\caption{Probability density of a subset of eigenstates from 
the lowest conduction band ($\nu=1\dots10$) for the spin-down electrons in a 
graphene SL of $N=20$ wells, $\Delta_{\rm ex}=8$ meV, 
$eV_{\rm SD}=\unit[5.5]{meV}$, $w_a=\unit[23.9]{nm}$ and $w_b=\unit[83.0]{nm}$. 
The baseline indicates the energy of every eigenstate. The right edge of the 
plot coincides with the edge of the SL.}
\label{WSL}
\end{figure}

In order to explore BOs in the tilted graphene SL, 
we consider a system of $N \gg 1$ wells with a source-drain voltage $V_{\rm SD}$ applied across the whole sample. 
Then, the energy 
spectrum of the graphene SL resembles the well-known 
Wannier-Stark ladder, as shown in Fig.~\ref{WSL}. This means that the 
energy levels ${E_\nu}$ become equally spaced with level spacing 
$eV_{\rm SD}/N$ and the eigenstates become localized with a similar 
envelope function. Only the states at the very edge of the energy 
spectrum (see, e.g., state labeled $\nu=1$ in Fig.~\ref{WSL}) are
influenced by finite size effects since they are localized close to the
boundaries of the SL. In view of this energy spectrum, BOs
are expected to occur in the device. 

\begin{figure}[tb]
\includegraphics[width=\linewidth,clip]{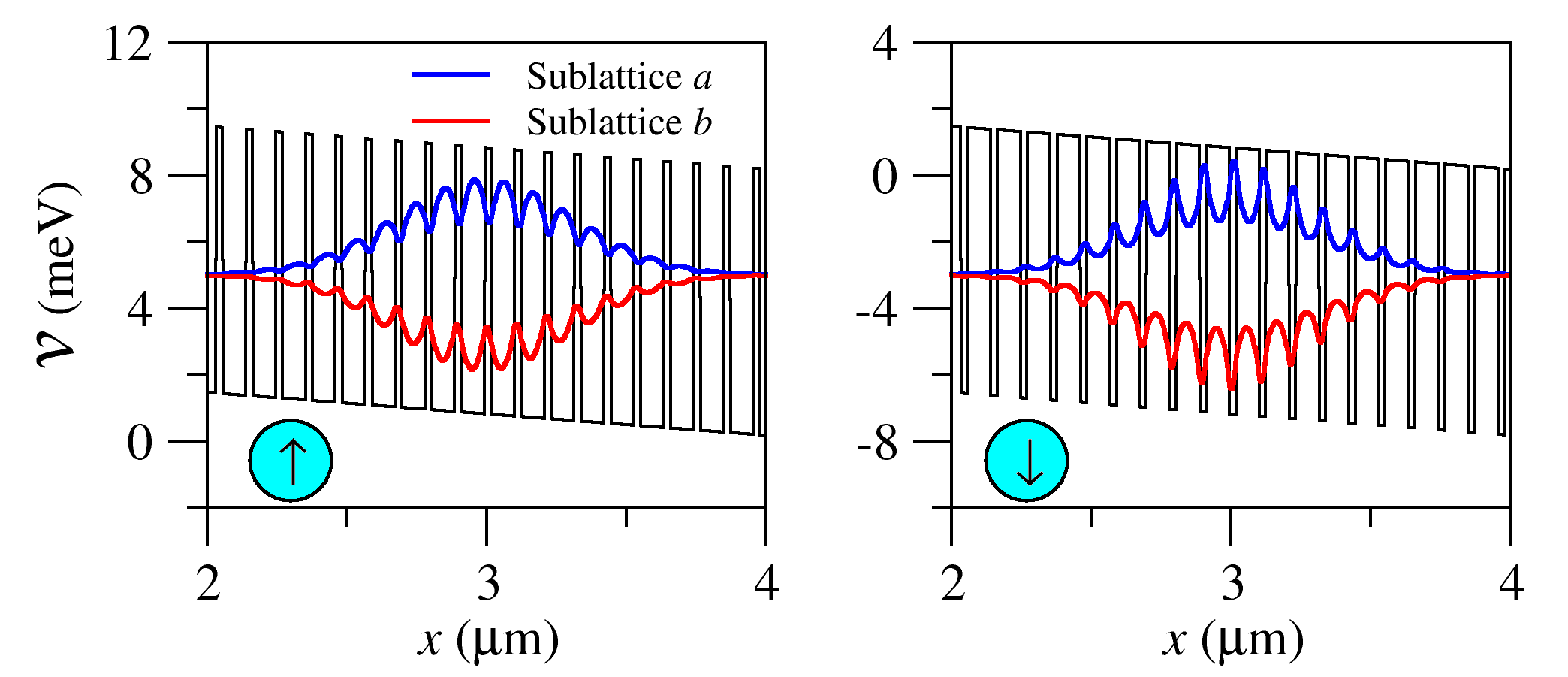}
\caption{\label{figInitial}
Potential profile and the initial state (arbitrary units) for 
spin-up (left) and spin-down (right) electrons in the biased device. 
Blue and red lines correspond to $f_a$ and $f_b$, respectively. 
The parameters are the same as in 
Fig.\ \ref{WSL}, but $N=80$ wells, $\sigma=3w=\unit[320.7]{nm}$ and 
$x_0=\unit[3.0]{\mu m}$.}
\end{figure}

As initial state we consider the localized wave packet of an electron 
excited to the lowest conduction band of the SL. Thus, we take 
the state $k=0$, as marked in Fig.~\ref{figBandstructure}(b) and (c) by 
red dots, multiply it with a Gaussian envelope of variance $\sigma^2$ 
centered at $x_0$ and normalize the wave function afterwards. In 
Fig.~\ref{figInitial}, the initial states for the 
spin-up and spin-down electrons are shown in a biased lattice of $N=80$ 
wells. Hereafter the initial states considered in the numerical 
simulations will be defined with parameters $\sigma=3w=\unit[320.7]{nm}$ and 
$x_0=\unit[3.0]{\mu m}$.

\begin{figure}[tb]
\includegraphics[width=0.8\linewidth,clip]{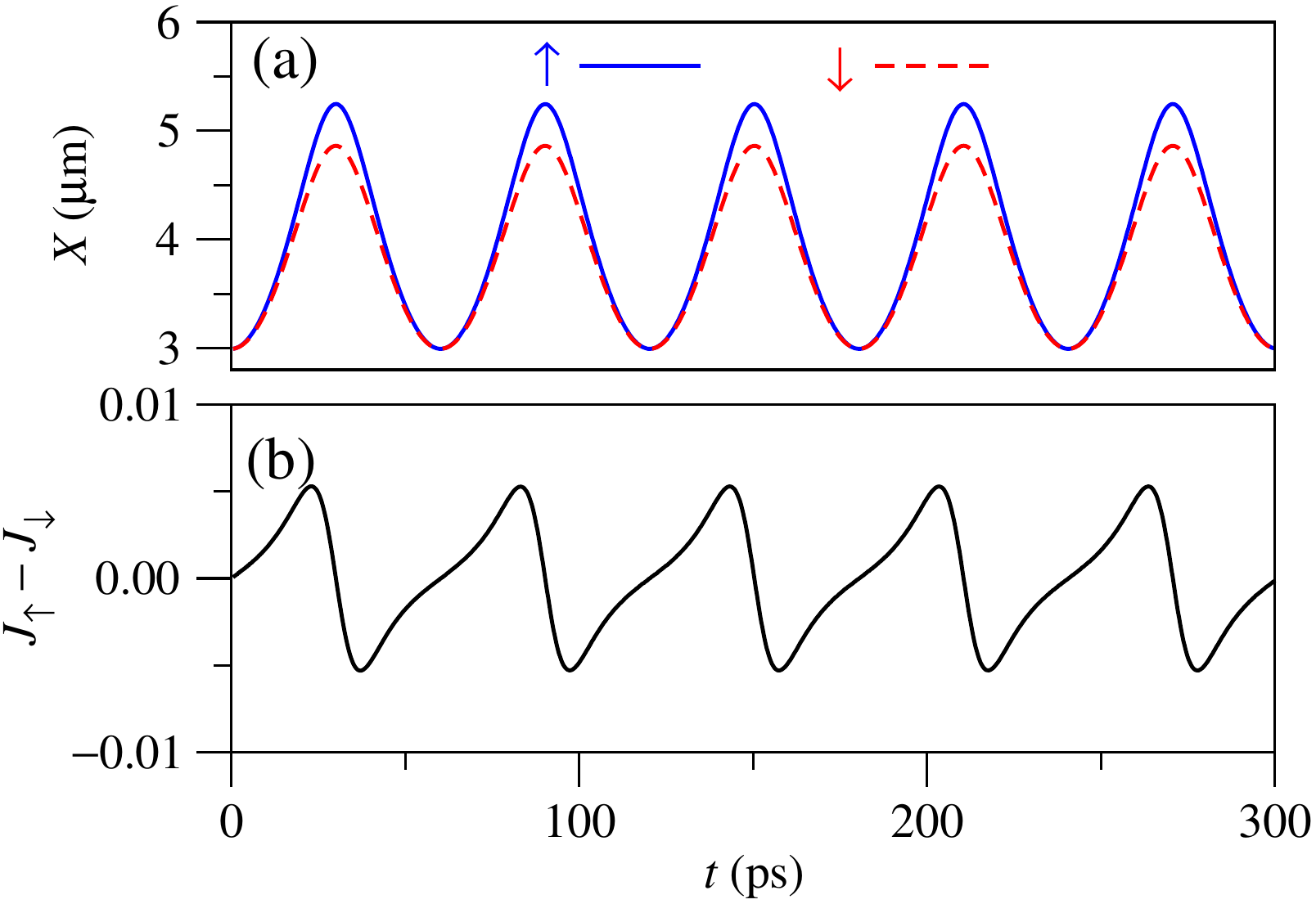}
\caption{\label{CC}
Time evolution of (a)~the centroid of spin-up
and spin-down electrons and (b)~the net polarized current in the 
biased device. The current data has been downsampled to a time resolution of 
$\unit[1]{ps}$. Parameters are the same as in 
Fig.~\ref{figInitial}.}
\end{figure}

The time-dependent wave function is obtained from the expansion of the initial 
wave function in the eigenstates $g_j^\nu$ of the tilted system as follows
\begin{equation}
\label{timeevolution}
g_j(t) = \sum_\nu c_\nu\,g_j^{\nu}\, e^{-i E_\nu t/\hbar}\ ,
\quad
c_\nu=\sum_j \bigl(g_j^{\nu}\bigr)^* g_j(0)\ .
\end{equation}
We study the electron dynamics by means of the time evolution of the
centroid of the wave function 
\begin{equation}
X(t)=d \sum_{j} j |g_{j}(t)|^2\ .
\label{centroid}
\end{equation}
We also define the dimensionless
current $J_s(t) \propto \langle \sigma_x \rangle$, where $\sigma_x$ is the 
Pauli matrix and $s$ refers to the spin, as
\begin{equation}\label{current}
J_s(t)=-\,\frac{\rmi}{2}\sum_{i,j} g_i(t)^* 
\big(\delta_{i,j+1}-\delta_{i,j-1}\big) g_j(t)\ .
\end{equation}
The electric current is proportional to the dimensionless current
$J_s(t)$. In Fig.~\ref{CC}, the time evolution of the centroid for 
spin-up and spin-down electrons and the net polarized current $J_\uparrow-J_\downarrow$ in a biased device are shown. 
Both magnitudes are clearly oscillatory with a well defined frequency of 
$\omega\approx\unit[0.1]{THz}$, which agrees well with the semiclassical 
estimate of the Bloch frequency $\omega_{\rm B}=eV_{\rm SD}/N\hbar$. 

\begin{figure}[bt]
\includegraphics[width=0.85\linewidth,clip]{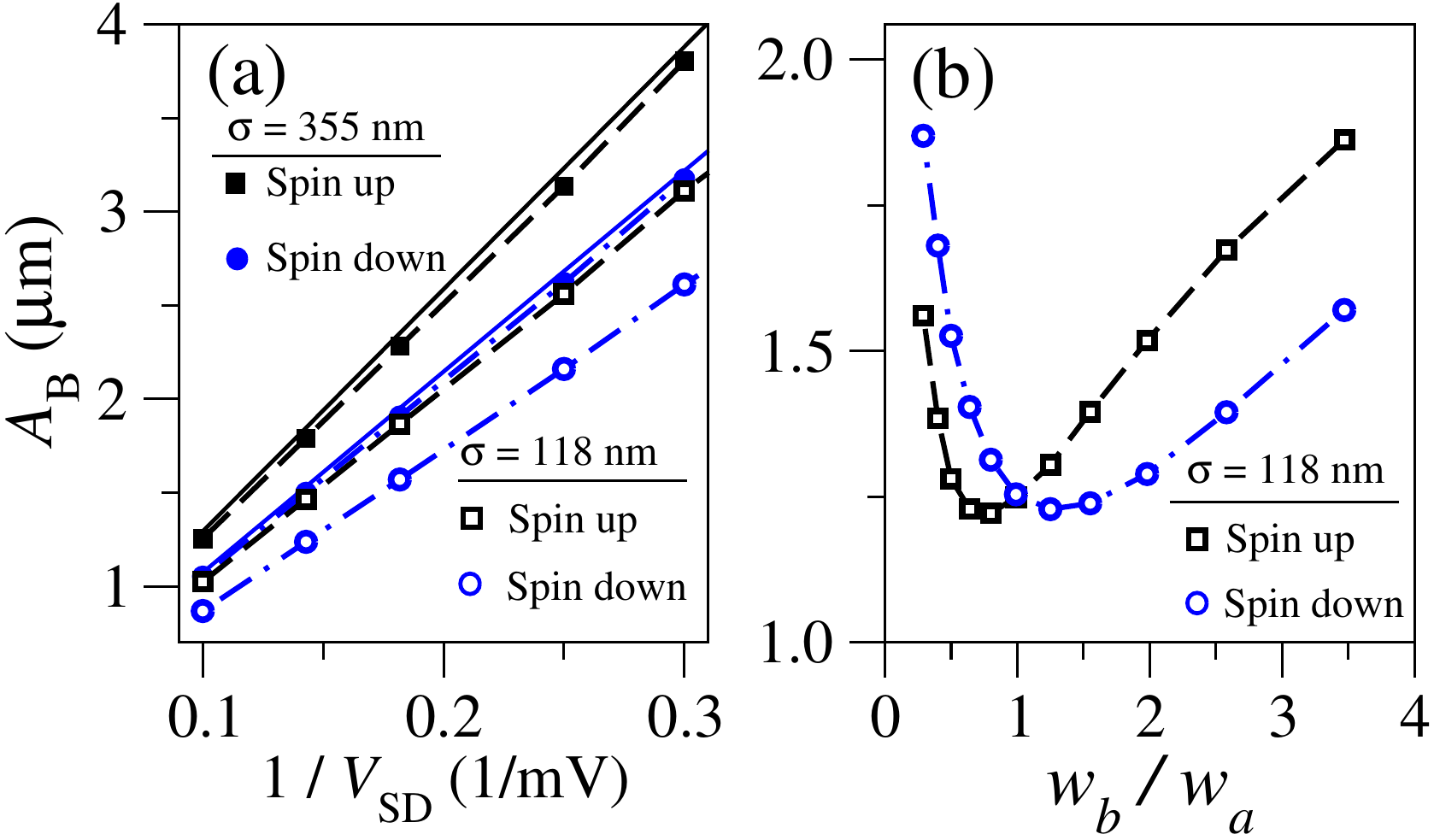}
\caption{(a)~Bloch amplitude of the spin-up and spin-down 
electrons as a function of the inverse voltage $1/V_{\rm SD}$.
Solid lines indicate the semiclassical prediction 
(infinite-width wave packet). (b)~Bloch amplitude 
as a function of the ratio between the width of the 
ferromagnetic strips and their spacing, with the constraint
$w_a+w_b=\unit[106.9]{nm}$. The other parameters are the same as in 
Fig.~\ref{figInitial}.}
\label{AB}
\end{figure}

Figure~\ref{AB}(a) shows the amplitude $\AB$ of the centroid motion as a 
function of the inverse of $V_{\rm SD}$ for both spins and two different 
values of $\sigma$.
Notice that the wider the wave packet, the better the amplitude approaches 
the classical estimate $\AB^{\uparrow \downarrow} = 
B_{\uparrow\downarrow}w/\hbar\omega_{\rm B}$, where $B_{\uparrow \downarrow}$ 
is the band width of the lowest conduction band (see Fig.~\ref{figBandstructure}).
Our simulations 
recover the expected dependence $\AB\sim\omega_{\rm B}^{-1}\sim V_{\rm SD}^{-1}$.
More importantly, although electrons
with spin up and spin down perform BOs with the same frequency $\omega_B$, 
the amplitude depends on the particular spin state. 

In Fig.~\ref{AB}(b), we study the impact of the geometry of the device 
on the spin selectivity by plotting the amplitude of the BOs as a function of the ratio of 
the spacing between the strips and the width of the strips. 
No spin-dependent effect is expected if the widths of 
both materials are the same. On the contrary if one of the widths 
is much larger than the other, the difference between the $\AB$ of both spin 
states is increased. Indeed, it can reach a difference up to 17\thinspace\% 
within the considered range of parameters. 

In conclusion, we have proposed a new design of THz oscillator based on 
hybrid graphene SLs. A spin-dependent potential acts on the 
electrons due to a set of ferromagnetic insulator strips deposited on top 
of a GNR. When subjected to a potential drop 
between source and drain, the electrons excited to the lowest 
conduction band perform BOs in the THz range. The frequency of the 
coherent oscillation is independent of the electron spin. On 
the contrary, the Bloch amplitude may differ significantly due to the 
different bandwidths for both spins. The different spatial extent of the 
electron motion in real space yields a spin-polarized ac electric current 
in the THz domain. The resulting ultrafast magnetization could be detected
with THz emission spectroscopy.\cite{Beaurepaire04}

\vspace{1ex}

We thank E. Diez and Y. M. Meziani for helpful discussions. Work in
Madrid was supported by MINECO (projects MAT2010-17180 and 
MAT2013-46308).

\vspace{1ex}

\bibliography{references}

\end{document}